\documentclass[a4paper,12pt]{article}
\usepackage{float}
\usepackage{tabularx}
\usepackage{amsmath}
\usepackage{amsfonts}
\usepackage{adjustbox}
\usepackage{multirow}
\usepackage{color}
\usepackage{longtable}
\usepackage{pdflscape}
\usepackage{rotating}
\usepackage{xcolor}

\usepackage{xcolor}
\usepackage[colorlinks = true,
            linkcolor = blue,
            urlcolor  = blue,
            citecolor = blue,
            anchorcolor = blue]{hyperref}
\usepackage[margin=1in,footskip=0.2in]{geometry}
\title{Algorithm to Obtain Inverse Potentials for $\alpha-\alpha$ Scattering using Variable Phase Approach}
\author{Anil Khachi\,$^{1, 2*}$, Shikha Awasthi\,$^{2}$, Lalit Kumar\,$^{2}$ and O.S.K.S. Sastri$^{2}$\\\\
$^{1}$ Department of Physics\\ St. Bedes College, 171002, \\Himachal Pradesh, Bharat (India)\\
\\
$^{2}$ Department of Physics and Astronomical Sciences\\ Central University of Himachal Pradesh\\ Dharamshala, 176215,H.P., Bharat (India)}

\begin{document}
\maketitle
\abstract{\noindent An algorithm$^{\ref{Fig1}}$ has been developed with the purpose of obtaining inverse potentials, where the Riccati-type non-linear differential equation, also called phase equation, has been kept in tandem with the Variational Monte Carlo method. The optimization of Gaussian function parameters is achieved such that the experimental phase shifts are reproduced. The obtained SPS for various $\ ell$ channels has been compared with experimental ones with mean absolute percentage error (MAPE) as a measure. The model parameters have been optimised by suitable optimisation technique by looking for minimum value of MAPE. The results for $\ell$=0$^+$, 2$^+$ and 4$^+$ partial waves have been obtained, to match with experimental SPS, with MAPE values of $2.9$, $4.6$ and $6.2$ respectively for data up to $23$ MeV, while for higher states 6$^+$, 8$^+$ and 10$^+$ has MAPE of $3.2, 4.5$ and $5.9$ respectively for data from $53-120$ MeV. On extrapolation for data in range E$_{\ell ab.}$ = $23-120$ MeV, using the optimised parameters, the SPS are found to be in close agreement with experimental ones for the first three channels.}\\
\\
{\textbf{keywords:} Inverse potentials, alpha-alpha scattering, phase function method (PFM), scattering phase shifts, Double Gaussian potential}

\section{Introduction}
In the 1950s, the first writings about inverse problems emerged in fields like physics (quantum scattering theory, electrodynamics, and acoustics), geophysics (electro-, seismo-, and geomagnetic exploration), and astronomy. With the advent of powerful computers, these problems found use in almost every academic area that uses mathematical models, including medicine, industry, ecology, economics, linguistics, and social sciences. To understand the structure of the nucleus one of the essential methods involved is scattering phenomena. The core purpose of this paper is to investigate scattering of alpha particles which are in relative motion for all partial waves using an inverse approach. Rutherford and Chadwick were the first who experimentally studied $\alpha-\alpha$ scattering in the year 1927 and since then a large amount of experimental data is available given by (i) Afzal \textit{et al.} \cite{Afzal} (ii) S. Chien and Ronald E brown \cite{Chien} (iii) Igo \cite{Igo} (iv) Darriulat, Igo, Pug, Holm \cite{Darriulat}(v) Nilson \cite{Nilson} and others. Alpha-alpha problem has been extensively studied both experimentally and theoretically with alpha particle having some sole properties like (i) zero spin and isospin (ii) tight binding energy of $28.3$ MeV having property to form cluster-like states for lighter nuclei ($^{6,7}$Li, $^9$Be, $^{12}$C and $^{16}$O are $\alpha$-structured) with alpha particle being the core nuclei in the cluster (iii) small root mean square radius of $1.44$ fm.\\
In the 1940's for $\alpha-\alpha$ scattering experiments, only naturally occurring $\alpha$-sources like polonium, thorium and radium were used, which did not result in very accurate results from experiments. Later on, with the advancement of technology, accelerators were used in scattering processes and highly accurate phase shifts were observed. The importance of $\alpha-\alpha$ scattering is that the study provides information regarding the force field in the nearest surrounding of He-nuclei and also provides information regarding energy levels of $^8$Be nucleus.\\ 
Haéfner was one of the first who studied $^8$Be properties in 1951 by using phenomenological potential \cite{Hafner}. Later on, Nilson, Briggs, Jentschke and others used Haéfner potential for $\ell = $2 state from the ground state of $96$~KeV. Later on, Nilson \textit{et al.} \cite{Nilson} extended Hafner model \cite{Hafner} to include $\ell$= 0, 2 and 4 and found that best agreement of phase shift with experiments requires small value of $R = 3.49$ fm. Also, Nilson found that testing the efficiency of the potential required data above $22.9$ MeV, which was unfortunately not available at that time. Later in the year 1958 Spuy and Pienaar \cite{Van} made phenomenological analysis up to $6$ MeV where they concluded that for $E<6$ MeV one needed velocity-dependent interaction for fitting S and D waves. Later on, Wittern \cite{Wittern} in the year 1959 derived the same semi-phenomenological potential and reached to same conclusion as given by Spuy and Pienaar. Later from 1960-to 1965 more phenomenological study was done by Igo who made an optical model analysis for $\alpha-\alpha$ in range $E = 23.1-47.1$ MeV and Darriulat, Igo and Pugh \cite{Igo} who used energy independent but strongly $\ell$ dependent complex Wood Saxon potential for range $E = 53-120$~MeV where they failed to fit the phase shifts using one single common potential for all the partial waves. Thus, it has been concluded, that single potential common to all $\ell$ do not exist phenomenologically. In short, $\alpha-\alpha$ potentials are found to be strongly $\ell$ dependent.\\
%%%%%%%%%%%%%%%%%%%%%%%%%%%%%%%%%%%%%%%%%%%%%%%%%%%%%%%%%%%%%%%%%%%%%%
Buck \textit{et al.} \cite{Buck} used Gaussian potential of the form
 \begin{equation}
     V(r)=-V_a exp{(-\alpha r^2)}+z_1z_2\frac{e^2}{r}erf(\beta r)
 \end{equation}
having single term with only two free parameters V$_a$ ($-122.6225$~MeV) and $\alpha$ ($0.22$~$fm^{-2}$). The \textit{erf()} function was included to take into account the Coulomb interaction and model parameters were obtained to fit all even $\ell$ channels up to $E_{\ell ab.} <$ $80$~MeV. Ali and Bodmer \cite{Ali} proposed two-term phenomenological potential with 4 parameters and were able to fit the phase shift for $\ell$ = 0, 2 and 4 partial waves with good accuracy to the experimental data. Later Darriulat \textit{et al.}, \cite{Darriulat} used the following Woods-Saxon potential, which has more than 6 parameters:
 \begin{equation}
 V_{\alpha \alpha}(r)=\frac{U_1}{[1+exp(r-r_1)/a_1]}-\frac{U_2}{[1+exp(r-r_2)/a_2]}+i\frac{W}{[1+exp(r-r_3)/a_3]}+V_C(r)
\end{equation}
with inelastic effects duly included in it for laboratory energy $53-120$~MeV and $\ell$= 0, 2, 4, 6, 8 and 10 partial waves phase shift was fitted for $E > 53$ MeV.\\
Phase function method (PFM) was used by Jana \textit{et al.} \cite{Jana} for $\alpha-\alpha$ scattering by using angular momentum-dependent complex Saxon-Woods potential as was suggested by Darriulat et al for $\ell$=0,2 and 4 partial wave only. Odsuren \textit{et al.} \cite{Odsuren} calculated partial scattering cross section using Gaussian potential for $\alpha-\alpha$, $n-\alpha$ and $p-\alpha$ using complex scaling method (CSM). Scattering phase shifts are commonly obtained analytically using S-matrix \cite{Mackintosh} and Jost function methods \cite{Jost}. Recently, there has been a renewed interest in the use of the Phase function method (PFM) or Variable phase approach (VPA) by Laha and group and they have applied the method for studying various light nuclei scattering problems which include the study of nucleon-nucleon (N-N), nucleon-nucleus (N-n) and nucleus-nucleus (n-n) \cite{Laha,Khirali, Sahoo, Bhoi} scattering using a variety of two-term potentials such as modified Hulthen and Manning-Rosen. While traditional S-matrix approaches depend on wave functions obtained by solving time independent  Schr$\ddot{o}$dinger equation (TISE), PFM requires only potential function to obtain the scattering phase shifts. PFM has been successfully applied on various interactions like neutron-proton \cite{AnilPRC, Scripta}, proton-proton \cite{Anil}, neutron-deuteron \cite{PAN, JNP, JNP2} and $\alpha-\alpha$ \cite{Khachi} interactions.\\
In this paper, our main objective is to obtain the scattering phase-shifts for alpha-alpha system using two term Gaussian potential and Coulomb term included as an $erf()$ function in an elastic region only i.e ($E_\ell$=$0-23$~MeV) by employing PFM for all even partial waves including $\ell$ = 6, 8 and 10. We have included the above partial waves because although their contribution to the total cross-section is small, it can not be neglected. Above the elastic region, we have extrapolated the results up to $120$~MeV energies. Instead of following Ali procedure \cite{Ali} we have given free run to the parameters for all partial waves. We have extended the calculations for higher $\ell$ channels like $\ell$=6,8 and 10 which is missing in Ali \textit{et al.} work. PFM has been employed in tandem with the model parameter optimising technique to obtain the results.
%%%%%%%%%%%%%%%%%%%%%%%%%%%%%%%%%%%%%%%%%%%%%%%%%%%%%%%%%%%%%%%%%%%%%%%%%
\section{Methodology:}\label{sec2}
\subsection{Model of Interaction:} 
The interaction between the two alpha particles is written as a combination of nuclear and Coulomb part as 
\begin{equation}
V_{\alpha\alpha}= V_{\alpha\alpha}^{(N)}+V_C
\end{equation}
where the nuclear part is
\begin{equation}
V_{\alpha \alpha}^{(N)}(r)=V_{r} exp(-\mu_{r}^2 r^2)-V_{a} exp(-\mu_{a}^2 r^2)
\end{equation}
Here, V$_{r}$ is attraction strength and V$_{a}$ is repulsion in MeV. $\mu_{r}$ and $\mu_{a}$ are the inverse ranges in $fm^{-2}$. The Coulomb potential $V_c$ has form
\begin{equation}
V_{c}=\frac{4e^2}{r}erf(\beta r)~~  \text{and}~~\beta=\frac{\sqrt{3}}{2R_\alpha}
\end{equation}
Interaction from this expression is called improved Coulomb interaction.  It is due to the finite size of the $\alpha$-particles, which is given by RMS value of its radius. That is, R$_\alpha$  = 1.44 fm.
\subsection{Phase Function Method:} 
The Schr$\ddot{o}$dinger wave equation for a spinless particle with energy E and orbital angular momentum $\ell$ undergoing scattering with interaction potential V(r) is given by
\begin{equation}
\frac{\hbar^2}{2\mu}\bigg[\frac{d^2}{dr^2}+\bigg(k^2-\frac{\ell(\ell+1)}{r^2}\bigg)\bigg]u_\ell(k,r)=V(r)u_\ell(k,r)
\end{equation}
Where
\begin{equation}
k_{c.m}=\sqrt{\frac{2\mu E_{c.m}}{{\hbar^2}}}~fm^{-1}
\end{equation}
with $\frac{\hbar^2}{2\mu}=10.44217$ MeV $fm^2$. For $\alpha-\alpha$ system, center of mass energy $E_{c.m.}$ is related to laboratory energy by following the relation for non-relativistic kinematics 
\begin{equation}
E_{c.m}=\frac{M_\alpha}{M_\alpha+M_\alpha}E_{\ell ab}=0.5E_{\ell ab}
\end{equation}
PFM or VPA is one of the important tools in scattering studies for both local \cite{Sastri} and non-local interactions \cite{Jana}. The mathematical foundation of the PFM method is well known in the theory of differential equations, that a linear homogeneous equation of second order, such as Schr$\ddot{o}$dinger equation, can be reduced to a nonlinear differential equation (NDE) of first order-the Riccati equation \cite{Morse}. The phase equation which was independently worked out by Calogero \cite{Calogero} and Babikov \cite{Babikov} is written in the following form.
\begin{equation}
\delta_{\ell}'(k,r)=-\frac{V(r)}{k(\hbar^2/2\mu)}\big[\cos(\delta_\ell(r))\hat{j}_{\ell}(kr)-\sin(\delta_\ell(r))\hat{\eta}_{\ell}(kr)\big]^2
\label{PFMeqn}
\end{equation}
This NDE is numerically integrated from origin to the asymptotic region using suitable numerical techniques, thereby obtaining directly the values of scattering phase shift for different values of projectile energy in a laboratory frame. The central idea of VPA is to obtain the phase shift $\delta$ directly from physical quantities such as interaction potential V(r), instead of solving TISE for wave functions u(r), which in turn are used to determine $\delta_{\ell}(k,r)$. With initial condition $\delta(0)=0$. The phase shift $\delta_{\ell}$ can be seen as a real function of $k$ and characterizes the strength of scattering of any partial wave i.e. say $\ell^{th}$ partial wave of the potential V(r). In the above equation and are the Bessel functions. Since we are only focusing on obtaining scattering phase shifts for $\ell$= 0 partial wave, the Riccati-Bessel function \cite{Calogero} is given by $\hat{j_{0}}=\sin(kr)$ and similarly the Riccati-Neumann function is given by $\hat{\eta_{0}}=-\cos(kr)$, thus reducing eq. \ref{PFMeqn} to
\begin{align}
  \delta_0'(k,r) &= \begin{aligned}[t]
      &  -\frac{V(r)}{k(\hbar^2/2\mu)}\bigg[  \sin(kr+\delta_0)\bigg]^2 \\
%      &  
       \end{aligned}
\intertext{PFM equation for D-wave takes the following form}       
  \delta_2'(k,r) &= \begin{aligned}[t]
      &  -\frac{V(r)}{k(\hbar^2/2\mu)}\bigg[  -\sin{\left(kr+ \delta_2 \right)}-\frac{3 \cos{\left(\delta_2 + kr \right)}}{kr} + \frac{3 \sin{\left(\delta_2 + kr \right)}}{k^{2} r^{2}}\bigg]^2 \\
      &  
       \end{aligned}
\intertext{PFM equation for G-wave takes the following form}
  \delta_4'(k,r) &= \begin{aligned}[t]
  \begin{cases}
&-\frac{V(r)}{k(\hbar^2/2\mu)}\bigg[\sin{\left(\delta_4 + kr \right)} + \frac{10 \cos{\left(\delta_4 + kr \right)}}{kr}-\frac{45 \sin{\left(\delta_4 + kr \right)}}{k^{2} r^{2}}- \frac{105 \cos{\left(\delta_4 + kr \right)}}{k^{3} r^{3}}\\
&+ \frac{105 \sin{\left(\delta_4 + kr \right)}}{k^{4} r^{4}}\bigg]^2
      \end{cases}
       \end{aligned}\\ 
\intertext{PFM equation for $\ell$=6 is}
\delta_6'(k,r) &= \begin{aligned}[t]
& -\frac{V(r)}{k(\hbar^2/2\mu)}\bigg[\cos(\delta_6(k,r))\hat{\tau}(kr)-\sin(\delta_6(r))\hat{\xi}(kr)\bigg]^2
 \end{aligned}\\
 \intertext{ where $\hat{\tau}(kr)$ is calculated as }
\hat{\tau}(kr) &= \begin{aligned}[t]
\begin{cases}
&(kr)^6\bigg\{\bigg(-21(kr)\bigg[495-60(kr)^2+(kr)^4\bigg]\cos(kr)\\
&+\bigg[-10395+4725(kr)^2-210(kr)^4+(kr)^6\bigg]\sin(kr)    \bigg) \bigg\}
\end{cases}
 \end{aligned}\\
  \intertext{ and $\hat{\xi}(kr)$ is calculated as}
\hat{\xi}(kr) &= \begin{aligned}[t]
\begin{cases}
&(kr)^6\bigg\{\bigg(-21(kr)\bigg[495-60(kr)^2+(kr)^4\bigg]\sin(kr)\\
&+\bigg[-10395+4725(kr)^2-210(kr)^4+(kr)^6\bigg]\cos(kr)\bigg) \bigg\}
\end{cases}
 \end{aligned}\\
   \intertext{PFM equation for $\ell$=8 is}
\delta_8'(k,r) &= \begin{aligned}[t]
&-\frac{V(r)}{k(\hbar^2/2\mu)}\bigg[\cos(\delta_8(k,r))\hat{\zeta}(kr)-\sin(\delta_8(k,r))\hat{\chi}(kr)\bigg]^2
 \end{aligned}\\
    \intertext{where we have}
 \hat{\zeta}(kr) &= \begin{aligned}[t]
 \begin{cases}
&\frac{1}{(kr)^8}\bigg\{9(kr)\bigg[ -225225+30030(kr)^2-770(kr)^4+4(kr)^6 \bigg]\cos(kr)\\
&+\bigg[ 2027025-945945(kr)^2+51975(kr)^4-630(kr)^6+(kr)^8\bigg]\sin(kr)  \bigg\}
\end{cases}
 \end{aligned}\\
     \intertext{and $\hat{\chi}(kr)$ is calculated to be}
  \hat{\chi}(kr) &= \begin{aligned}[t]
  \begin{cases}
&\frac{1}{(kr)^8}\bigg\{9(kr)\bigg[ -225225+30030(kr)^2-770(kr)^4+4(kr)^6 \bigg]\sin(kr)\\
&-\bigg[ 2027025-945945(kr)^2+51975(kr)^4-630(kr)^6+(kr)^8\bigg]\cos(kr)  \bigg\} 
\end{cases}
 \end{aligned}\\
      \intertext{PFM equation for $\ell$=10 takes form}
\delta_{10}'(k,r) &= \begin{aligned}[t]
&-\frac{V(r)}{k(\hbar^2/2\mu)}\bigg[\cos(\delta_{10}(k,r))\hat{\kappa}(kr)-\sin(\delta_{10}(k,r))\hat{\gamma}(kr)\bigg]^2 
 \end{aligned}\\
\intertext{where $\hat{\kappa}(kr)$ is calculated to be}
\hat{\kappa}(kr) &= \begin{aligned}[t]
\begin{cases}
&-(55 (kr)\bigg[11904165-1670760 (kr)^2 + 51597 (kr)^4-4684 (kr)^6 \\
&+(kr)^8\bigg]
 \cos(kr)+ \bigg[-654729075 + 310134825 (kr)^2- 18918900 (kr)^4 \\&+ 
315315 (kr)^6-1485 (kr)^8 + (kr)^{10}\bigg] \sin(kr))(kr)^{-10} 
\end{cases}
 \end{aligned}\\
\intertext{and $\hat{\gamma}(kr)$ comes out to be}
\hat{\gamma}(kr) &= \begin{aligned}[t]
\begin{cases}
&-(55 (kr)\bigg[11904165-1670760 (kr)^2 + 51597 (kr)^4-4684 (kr)^6 \\
&+(kr)^8\bigg]
 \sin(kr)+ \bigg[-654729075 + 310134825 (kr)^2- 18918900 (kr)^4 \\&+ 
315315 (kr)^6-1485 (kr)^8 + (kr)^{10}\bigg] \cos(kr))(kr)^{-10} 
\end{cases}
 \end{aligned}
\end{align}
These NDE’s equations (Eq. 10-21) are numerically integrated from origin to the asymptotic region using RK-4/5 method, thereby obtaining directly the values of scattering phase shift for different values of projectile energy in the lab frame. The central idea of VPA is to obtain the phase shift $\delta$ directly from physical quantities such as interaction potential V(r), instead of solving TISE for wave functions $u(r)$, which in turn are used to determine $\delta$. 
\subsection{Optimisation of model parameters using Variational Monte-Carlo (VMC) method:} 
The Variational Monte Carlo (VMC) method is an advanced technique which combines two important approaches i.e. the random or probability study characteristic of Monte Carlo simulations and the stringent optimisation principles of variational methods to explore the configuration space of a system efficiently. In the VMC technique, the inbuilt property of Monte Carlo simulations i.e. randomness is utilised to get closer to the desired configuration by iteratively modifying the model parameters. In our procedure, we consider potential model parameters to vary iteratively with respect to experimental data. The algorithm steps and block diagram$^{\ref{Fig1}}$ to implement VMC in computer is given below:
\begin{enumerate}
\item \textbf{Initialisation:}\\
In this step, we select the initial values for model parameters of double Gaussian potential: $V_a$, $\mu_a$, $V_r$ and $\mu_r$ by considering initial guess from theoretically available or empirical data. 
\item \textbf{Solving PFM equation:}\\
In this step, the PFM equation Eq. \ref{PFMeqn} is solved numerically using the Runge Kutta method (RK-5 method in this work) to obtain simulated scattering phase shifts (SPS) and named it as $\delta_{old}$. The mean absolute percentage error for simulated SPS w.r.t experimental data is calculated and saved as $MAPE_{old}$.
\item \textbf{Monte Carlo step:}\\
In this step, a random number say, `$r$' is generated within an interval [-I, I] and then add the perturbation `$r$' to one out of all four parameters e.g $V_{a_{new}}$ $=V_a$ $+ r$.
\item \textbf{Recalculating SPS:}\\
The scattering phase shift is again calculated by considering a new set of perturbed parameters i.e. $V_{a_{new}}$, $\mu_a$, $V_r$ and $\mu_r$. The mean absolute percentage error is again calculated and saved as $MAPE_{new}$.
\item \textbf{Variational step:}\\
In this step, the condition i.e. $MAPE_{new}< MAPE_{old}$ is checked. If this condition is true, then the parameter $V_a$ is updated to $V_{a_{new}}$ otherwise the old value is retained.
\item \textbf{Iterative steps:}\\
Repeat steps 3,4 and 5 for all the model parameters to complete one iteration. After a particular number of iterations, the size of interval `$r$' is reduced to check if there is any further reduction in MAPE. The procedure is finished when MAPE reaches convergence i.e. when the MAPE ceases to change.
\end{enumerate}

\begin{figure*}[htp]
\centering
{\includegraphics[scale=1.2]{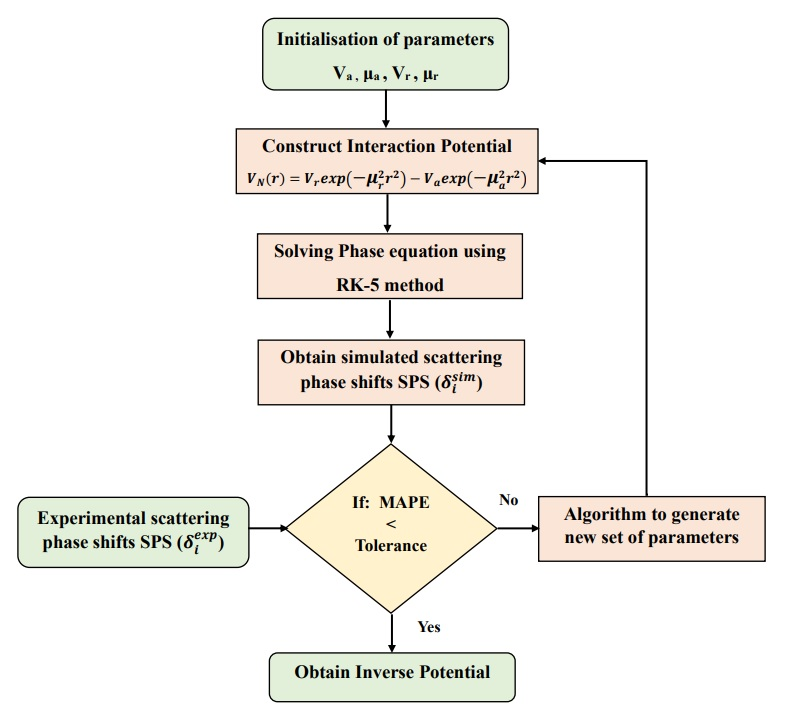}}
\caption{Block diagram to obtain inverse potentials for $\alpha-\alpha$ scattering.}
\label{Fig1}
\end{figure*}
%%%%%%%%%%%%%%%%%%%%%%%%%%%%%%%%%%%%%%%%%%%%%%%%%%%%%%%%
\section{Simulation of Results and Discussion}\label{sec3}
Here, we have used double Gaussian potential and calculated the SPS using the phase function method (PFM). Our obtained potentials are real and are obtained at elastic energies, $E_{\ell ab.}$= 0-23 MeV. The extrapolated curves are just an indication of how phase shift varies at higher energies (E $>$ 23 MeV) and are observed to be in good match with the experimental data up to 120 MeV on extrapolation. Some distortion from the experimental data above 23 MeV may be due to the absence of inelastic potential in our calculations. We have extended our work by taking additional $\ell$ = 6, 8 and 10 partial wave data which have not been investigated using PFM in any recent literature.\\
Potentials associated with partial waves are shown in figures \ref{fig3}. Standard procedures were used to numerically calculate the phase shifts using equations (10-19). Our approach has been to vary all 4 parameters for $\ell$  = 0-10, but we observed that for  $\ell$ = 0, 2 and 4 all parameters are required to get their respective phase shifts but for $\ell$ = 6, 8 and 10, it was observed that keeping or removing the repulsive core did not affect the phase shifts i.e. MAPE almost remained same, this must be due to the centrifugal barrier concealing the inner repulsive core. So, only two parameters were adjusted by to get phase shifts for $\ell$ = 6, 8 and 10 partial waves. A similar observation was found in work done recently by Darriulat \cite{Darriulat} and Laha \textit{et al.} \cite{Laha} where it was observed that a single static potential can not give phase shift for all the partial waves. Double-term Gaussian potential is giving similar nature potential as was given by Darriulat et al. with a repulsive core and an attractive outer region. At $E_{\ell ab.}$ = 0.6 MeV the value of our phase shift for $\ell$ = 0 is 182.2 degree while experimental value is 178 $\pm$ 1 degree. Phase shift for $\ell$ = 0 changes its sign at $E_{\ell ab.}$ = 20 MeV, which is in close agreement with the work done by Ali and Bodmer \cite{Ali}, Tombrello and Senhouse \cite{Tombrello} and Afzal \cite{Afzal} and continues to be in negative phase up to highest beam energy used ($120$~MeV). Our S-wave phase shift are consistent with experimental data (with MAPE=0.86$\%$) above $23$~MeV even on extrapolation and can be clearly observed in Figure 2. D-wave phase shift becomes appreciable at $E_\alpha$ $>2.5$ MeV and at $E_{\ell ab}$=11.8 MeV maximum value of our phase shift for $\ell$ = 2 is 115.56 degree while experimental value is 114.9$\pm$2 degree, while the G-wave shift rises from $E_\alpha$ $>11$ MeV and at $E_{\ell ab}$=77.5 MeV maximum value of our phase shift for G-wave is $140.52$ degree while experimental value is 137 $\pm$ 1.8 degree. The $\ell$=6 and $\ell$=8 phase shifts can be seen to become active above $E_\alpha$ $\approx$ $30$ and $50$~MeV respectively.\\
The repulsive core for $\ell$=0 is more strong than D, G, I, K and M partial waves which can be observed from Figure 3 and thus our potential is consistent with the results of Darriulat and Laha et al. Here it should be noted that the attractive interaction between two alpha particles is an important element which is responsible for keeping the nuclei against Coulombic repulsion. Also here it should be noted that the potential for $\ell$=6,8 and 10 in Figure 3 should not be taken too seriously since experimental data above $54$~MeV was only considered to get the nature of interaction so it must be dealt with cautiously. By utilising the PFM method we have been able to construct appropriate phenomenological potential for the entire range of energy for $\ell$=0-10 partial waves without incorporating the effects of inelastic processes. Including an inelastic process may further bring down the MAPE and may even give approximate realistic potentials for the entire energy range.
%%%%%%%%%%%%%%%%%%%%%%%%%%%%%%%%%%%%%%%%%%%%%%%%%%%%%%%%%%%%%%%%%%%%%%%%%%
\begin{table}[]
\centering
\begin{tabular}{c|c|c|c|c|c}
\hline
States & \begin{tabular}[c]{@{}c@{}}$V_a$\\ \textit{(MeV)}\end{tabular} & \begin{tabular}[c]{@{}c@{}}$\mu_a$\\$(fm^{-2})$\end{tabular} & \begin{tabular}[c]{@{}c@{}}$V_r$\\ \textit{(MeV)}\end{tabular} & \begin{tabular}[c]{@{}c@{}}$\mu_r$\\ $(fm^{-2})$\end{tabular} & \begin{tabular}[c]{@{}c@{}}MAPE\\ (\%)\end{tabular} \\ \hline
$S^+$     &   15.2767         & 0.1851        & 693.912          & 0.942        & 7.6                                                 \\ 
$D^+$     & 73.9102           & 0.4741        & 334.852          & 1.0441        & 2.7                                                                           \\ 
$G^+$     & 246.917           & 0.47458         & 235.731           & 3.9989                                                                                         & 6.5 \\ 
$I^+$     & 184.747           & 0.63178        & --               & --                                                & 1.7                                                 \\
$K^+$     & 6.28974            & 0.2271        & --               & --                                                  & 7.4                                                 \\ 
$M^+$     & 12.2422            & 0.25907        & --               & --                                                  & 10.4                                                 \\ 
\hline
\end{tabular}
\end{table}  
%%%%%%%%%%%%%%%%%%%%%%%%%%%%%%%%%%%%%%%%%%%%%%%%%%%%%%%%%%%%%%%%%%%
\begin{figure*}[htp]
\centering
{\includegraphics[scale=0.6]{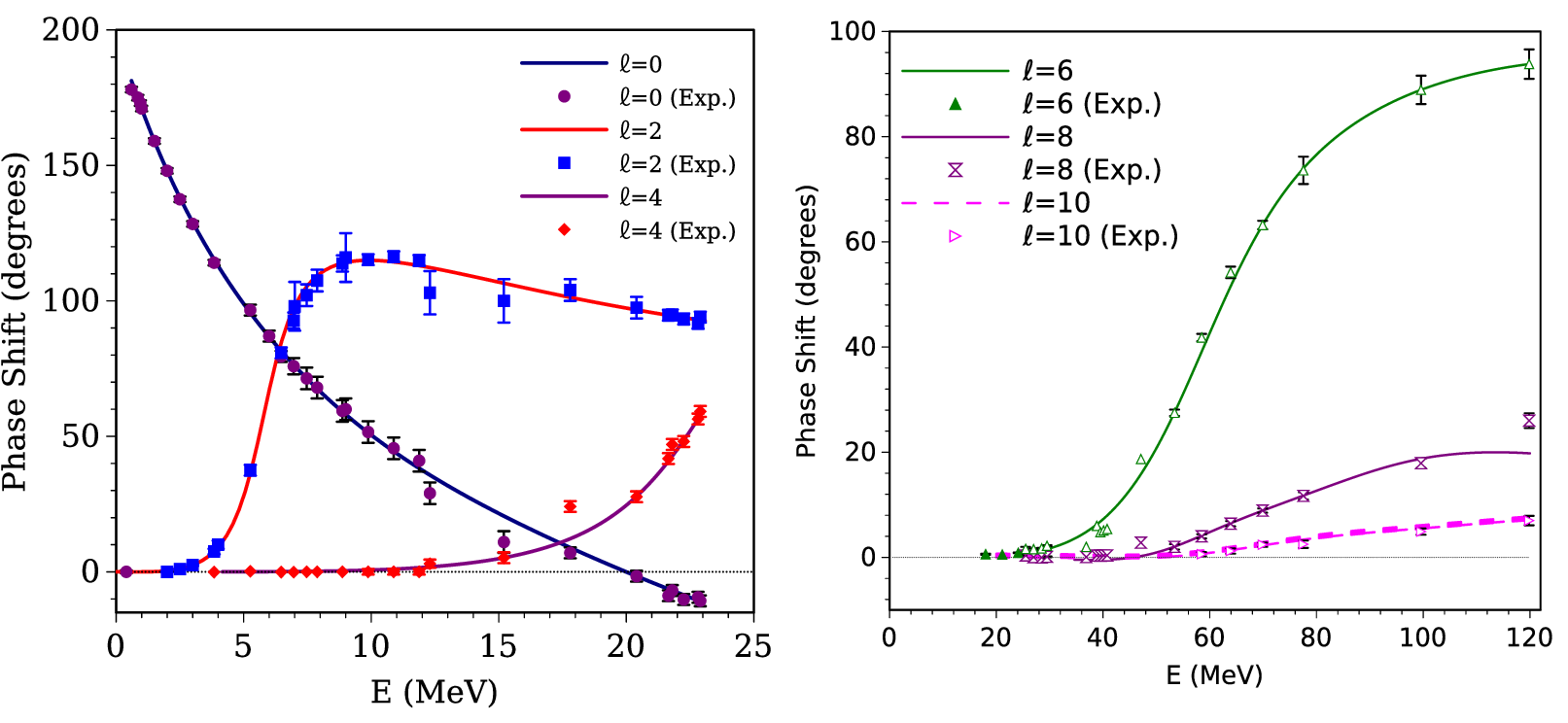}}
\caption{Phase shift for S, D, G (Left) and I,K, M (Right) states. Experimental data is taken from Heydenberg and Temmer, Russel, Phillips and Reich and Jones, Phillips and Miller E=0.15-9 MeV (1956-60), Nilson, Jentschke, Briggs, Kerman, Snyder E=12.3-22.9 MeV (1958).}
\label{fig1}
\end{figure*}
%%%%%%%%%%%%%%%%%%%%%%%%%%%%%%%%%%%%%%%%%%%%%%%%%%%%%%%%%%%%%%%%%%
\begin{figure*}[htp]
\centering
{\includegraphics[scale=1]{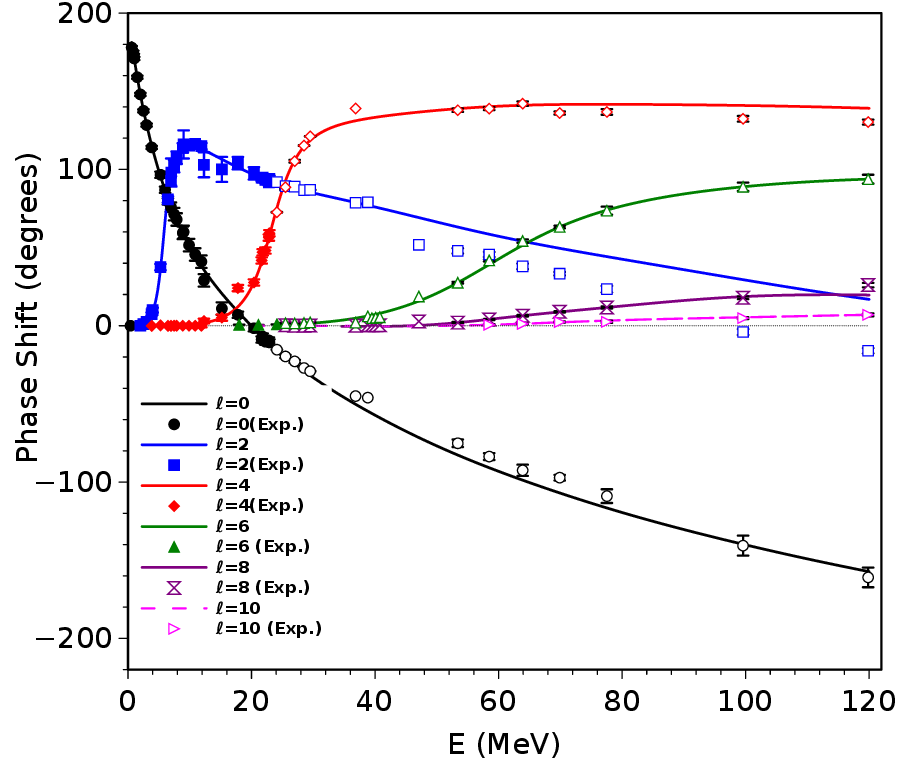}}
\caption{Phase shift for S, D, G, I, K and M-states. From 23 MeV onwards we have extrapolated data up to 120 MeV only for the first three channels. Experimental data is taken from Heydenberg and Temmer, Russel, Phillips and Reich and Jones, Phillips and Miller E=0.15-9 MeV (1956-60), Nilson, Jentschke, Briggs, Kerman, Snyder E=12.3-22.9 MeV (1958), Igo E=23.1-47.1 MeV (1960),  Tombrello and Senhouse E=3.84-11.88 MeV (1963), Darriulat, Igo and Pugh E=53-120 MeV (1965) and Chen and Ronald E=18-29.5 MeV (1974).}
\label{fig2}
\end{figure*}
%%%%%%%%%%%%%%%%%%%%%%%%%%%%%%%%%%%%%%%%%%%%%%%%%%%%%%%%%%%%%%%%%%
\begin{figure*}[htp]
\centering
{\includegraphics[scale=0.6]{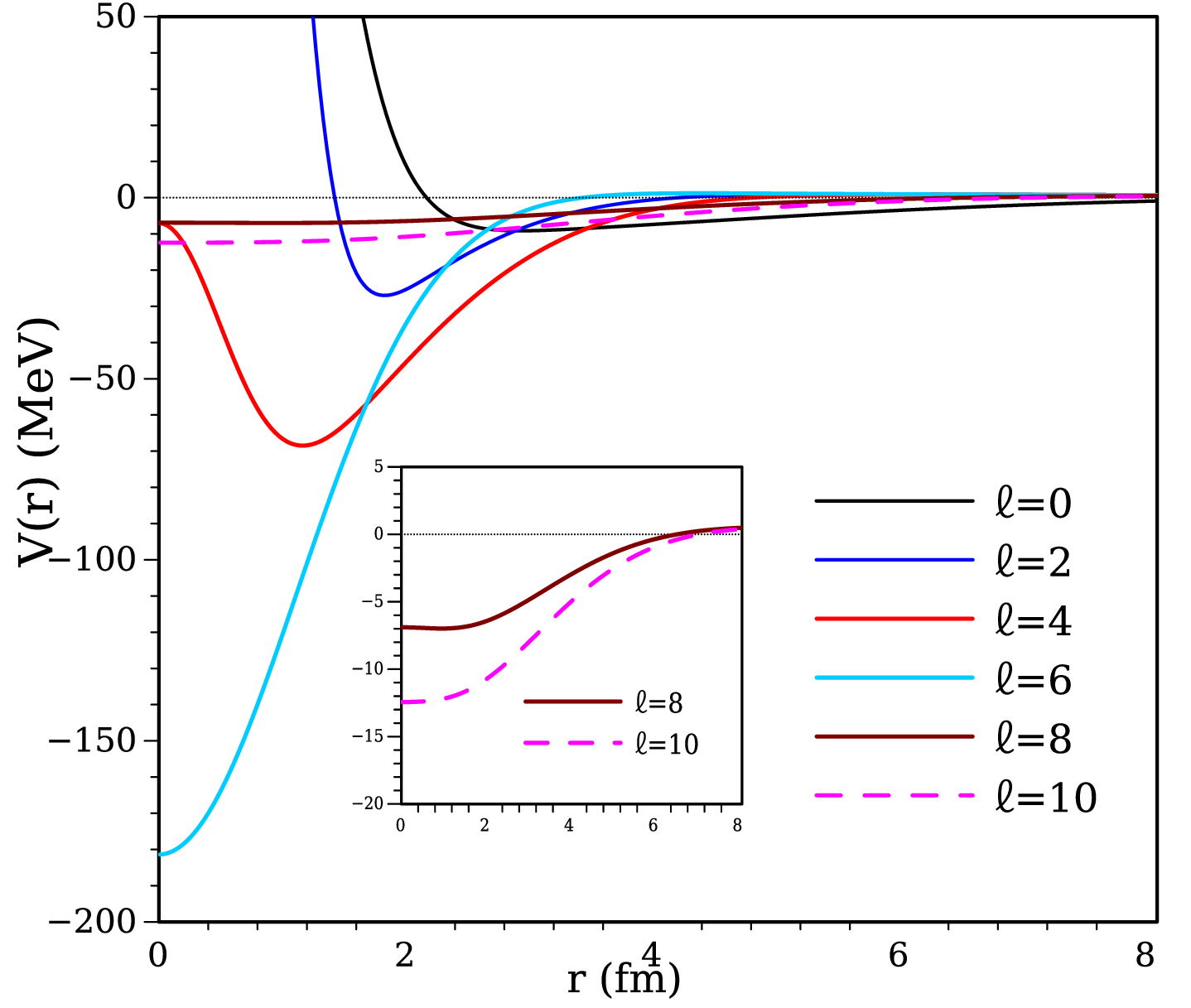}}
\caption{Potential plots for S, D, G, I, K and M-states}
\label{fig3}
\end{figure*}
%%%%%%%%%%%%%%%%%%%%%%%%%%%%%%%%%%%%%%%%%%%%%%%%%%%%%%%%%%%%%%%%%%%
\subsection{Phase shift, amplitude and wavefunction vs. distance $r$(fm)}
The equations and figures for phase shift, amplitude and wavefunction for $\ell$=0, 2 and 4 states are given below:
\subsubsection{Phase shift vs. distance $r$(fm)}
The PFM allows to observe the variation of phase shifts w.r.t the distance $r$. For $\ell$=0, 2 and 4 states, the phase shift vs. $r$(fm) plots are portrayed in figure \ref{fig5}.
%\begin{figure*}[htp]
%\centering
%{\includegraphics[scale=0.22]{delta0.png}}
%{\includegraphics[scale=0.2]{delta2.png}}
%{\includegraphics[scale=0.2]{delta4.png}}
%\caption{Phase shift vs $r$(fm) plots for $\ell$=0, 2 and 4 states.}
%\label{fig4}
%\end{figure*}
%%%%%%%%%%%%%%%%%%%%%%%%%%%%%%%%%%%%%%%%%%%%%%%%%%%%%%%%%%%%%%%%%%
\subsubsection{Amplitude vs. distance $r$(fm)}
\begin{figure*}[htp]
\centering
{\includegraphics[scale=0.6]{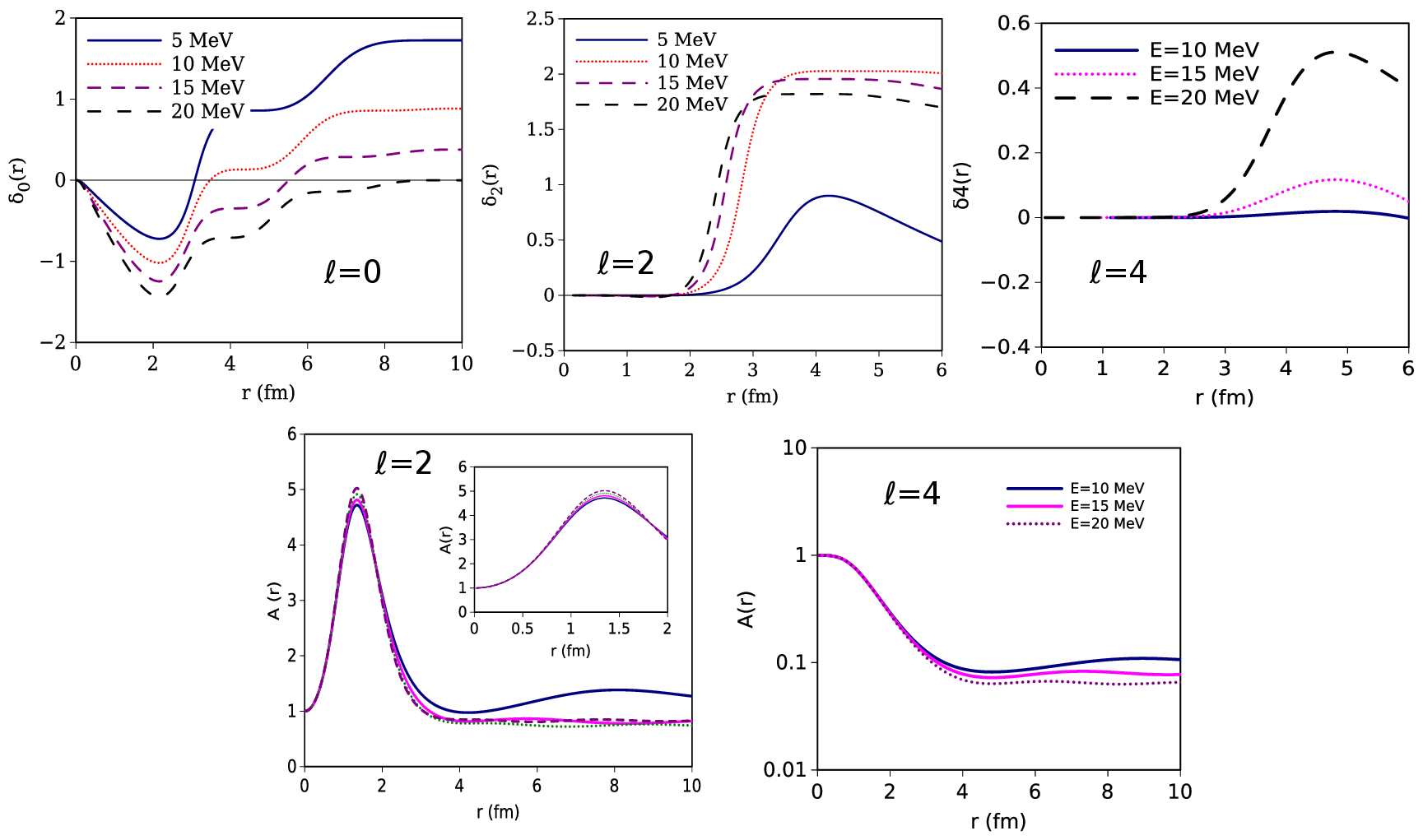}}
\caption{Phase shift vs $r$(fm) plots for $\ell$=0, 2 and 4 states and amplitude vs $r$(fm) plots for $\ell$= 2 and 4 states.}
\label{fig5}
\end{figure*}

The equation and figures for amplitude vs. $r$(fm) for $\ell$=0, 2 and 4 states are given in figures \ref{fig5}.

\begin{equation}
\begin{aligned}
A_{\ell}^{\prime}(r)= & -\frac{A_{\ell} V(r)}{k}\left[\cos \left(\delta_{\ell}(k, r)\right) \hat{j}_{\ell}(k r)-\sin \left(\delta_{\ell}(k, r)\right) \hat{\eta}_{\ell}(k r)\right] \\
& \times\left[\sin \left(\delta_{\ell}(k, r)\right)\left(\hat{j}_{\ell}(k r)+\cos \left(\delta_{\ell}(k, r)\right) \hat{\eta}_{\ell}(k r)\right]\right.
\end{aligned}
\end{equation}

\begin{align}
\intertext{Amplitude function equation for $\ell=0$ is given as}
A_0^{\prime} = & \begin{aligned}[t]
&-\frac{A_0 V(r)}{k\left(\frac{\hbar^2}{2 \mu}\right)}\left[\cos \delta_0 \sin (k r)-\sin \delta_0 (-\cos (k r))\right] \\
& \times\left[\sin \delta_0 \sin (k r)+\cos \delta_0 (-\cos (k r))\right]
\end{aligned}
\intertext{Amplitude function equation for $\ell=2$ is given as} 
A_2^{\prime}= & \begin{aligned}[t]
& -\frac{A_2 V(r)}{k\left(\frac{\hbar^2}{2 \mu}\right)}\left[\cos \delta_2\left(\left(\frac{3}{(k r)^2}-1\right) \sin (k r)-\frac{3}{(k r)} \cos (k r)\right)\right. \\
& \times\left(\sin \delta_2\left(\left(\frac{3}{(k r)^2}-1\right) \sin (k r)-\frac{3}{(k r)} \cos (k r)\right)+\cos \delta_2\left(\left(-\frac{3}{(k r)^2}+1\right) \cos (k r)\right.\right. \\
& \left.\left.-\frac{3}{(k r)} \sin (k r)\right)\right]
\end{aligned}
\end{align}
Similarly one can obtain an amplitude function equation for $\ell=4$ using suitable Bessel functions.

%----------------------------------------------
%%%%%%%%%%%%%%%%%%%%%%%%%%%%%%%%%%%%%%%%%%%%%%%%%%%%%%%%%%%%%%%%%%
\subsubsection{Wavefunction vs. distance $r$(fm)}
The equation and figures for wavefunction vs. $r$(fm) for $\ell$=0, 2 and 4 states are given in figures \ref{fig5}.
\begin{equation}
u_{\ell}(r)=A_{\ell}(r)\left[\cos \left(\delta_{\ell}(k, r)\right) \hat{j}_{\ell}(k r)-\sin \left(\delta_{\ell}(k, r)\right) \hat{\eta}_{\ell}(k r)\right]
\end{equation}

\begin{align}
\intertext{Wavefunction equation for $\ell=0$ is given as}
u_0(r)= & \begin{aligned}[t] A_0(r)\left[\cos \delta_0(r) \cdot \sin (k r)-\sin \delta_0(r) \cdot \cos (k r)\right]\\
\end{aligned}
\intertext{Wavefunction equation for $\ell=2$ is given as} 
u_2(r) = & \begin{aligned}[t]  
&A_2(r)\left[\cos \delta_2(r)\left(\left(\frac{3}{(k r)^2}-1\right) \sin (k r)-\frac{3}{(k r)} \cos (k r)\right)\right. \\
& \left.-\sin \delta_2(r)\left(-\left(-\frac{3}{(k r)^2}+1\right) \cos (k r)-\frac{3}{(k r)} \sin (k r)\right)\right]
\end{aligned}
\end{align}
Similarly one can obtain wavefunction equation for $\ell=4$ using suitable Bessel functions. 

\begin{figure*}[htp]
\centering
{\includegraphics[scale=0.6]{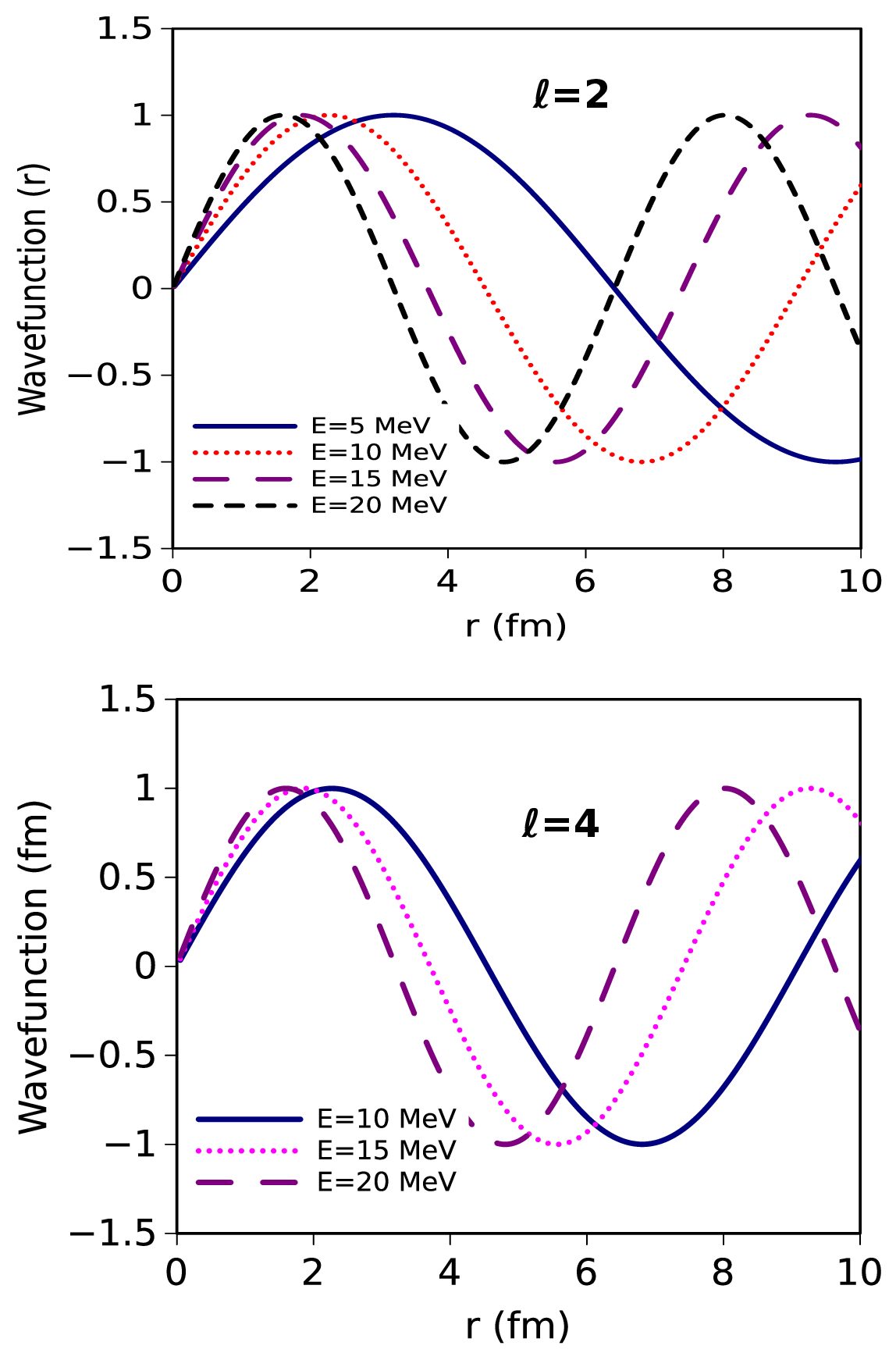}}
\caption{Wavefunction vs $r$(fm) plots for $\ell$= 2 and 4 states.}
\label{fig6}
\end{figure*}
%%%%%%%%%%%%%%%%%%%%%%%%%%%%%%%%%%%%%%%%%%%%%%%%%%%%%%%%%%%%%%%%%%
\subsection{Cross-section:}
Partial cross-section plot is shown in Figure 2 for different states calculated using the obtained SPS values. Partial cross-section has been calculated by expression \cite{Khachi} 
\begin{equation}
\sigma_{\ell}=\frac{4\pi}{k^2}(2\ell+1)\sin^2{\delta_{\ell}(k)}
\end{equation} 
%-------------------------------------------------------------------
\begin{figure*}[htp]
\centering
{\includegraphics[scale=0.6]{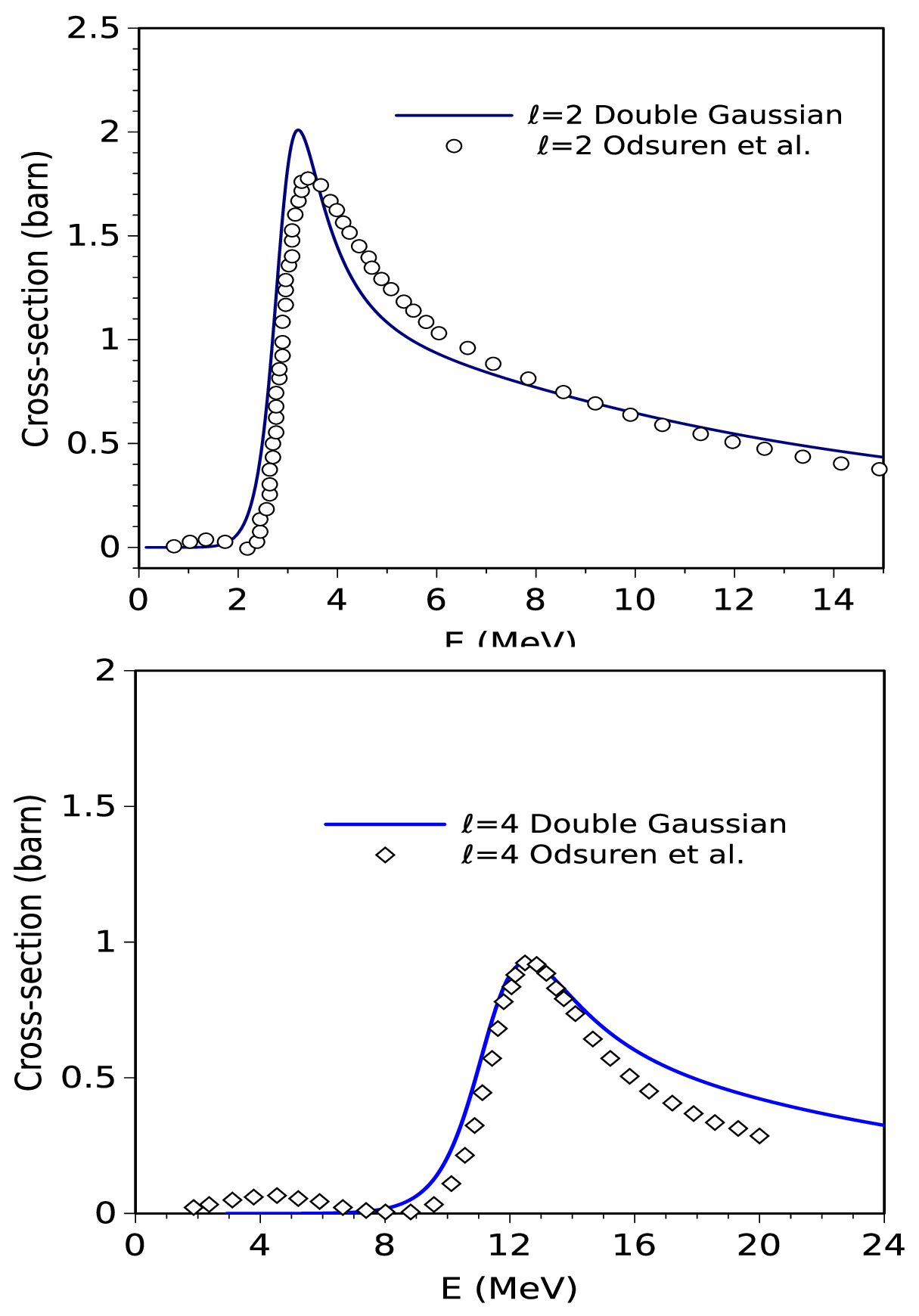}}
\caption{Cross section for D and G channels as a function of center of mass energy $E_{C.M}$.}
\label{fig7}
\end{figure*}
%----------------------------------
It can be observed that the wavefunction, amplitude and partial cross-section plots for partial wave $\ell=2,4$ are only given. Since for $\ell=0$ partial wave, the decay width is very narrow and sharp but the contribution of $\ell=2,4$ in cross section is rather significant due to broad resonance peak \cite{Odsuren}.
%%%%%%%%%%%%%%%%%%%%%%%%%%%%%%%%%%%%%%%%%%%%%%%%%%%%%%%%%%%%%%%%%%
\section{Conclusion}\label{sec4}
The scattering phase shifts for $\ell$ = 0 (S-channel), $\ell$ = 2 (D-channel) and $\ell$ = 4 (G-channel) have been computed up to 23 MeV and the best-fitted parameters are found to give a good match with the experimental data when extrapolated to the inelastic region of E $>$ 40 MeV up to $120$ MeV. For $\ell$ = 6 (I-channel), $\ell$ = 8 (K-channel) and $\ell$ = 10 (M-channel) also, a good match has been seen with the experimental data. Including an inelastic process may further bring down the MAPE and may even give approximate realistic potentials for the entire energy range and shall be taken up in future. At last, the computed cross-section results are compared with that of Odsuren \textit{et al.} and are found to be in good agreement. In addition to earlier works done by Laha \textit{et al.}, using PFM for $\ell$ = 0, 2 and 4, we have explored the applicability of PFM to even higher states up to $\ell$ = 6, 8 and 10. Thus, PFM stands as an efficient tool for phase shift calculations in quantum mechanical scattering problems for local as well as non-local potentials. We may summarise our entire work in 3 main points: \\
(i) Double Gaussian potential results in effective inverse interaction potentials for alpha-alpha scattering for all $\ell$-channels.\\
(ii) The Phase Function Method is an effective tool for phase shift calculations. \\
(iii) Phases shifts calculated for the elastic region have been found to give good results for inelastic energy data up to $120$ MeV for all partial waves. 

%\bmhead{Acknowledgments}
%
%Acknowledgments are not compulsory. Where included they should be brief. Grant or contribution numbers may be acknowledged.
%
%Please refer to Journal-level guidance for any specific requirements.

\section*{Declarations}
%
%Some journals require declarations to be submitted in a standardised format. Please check the Instructions for Authors of the journal to which you are submitting to see if you need to complete this section. If yes, your manuscript must contain the following sections under the heading `Declarations':

\begin{itemize}
\item Funding: No funding is approved for this work
\item Conflict of interest: There is no conflict of interest whatsoever.
%\item Ethics approval 
%\item Consent for publication
%\item Availability of data and materials
%\item Authors' contributions
\end{itemize}

%\noindent
%If any of the sections are not relevant to your manuscript, please include the heading and write `Not applicable' for that section. 




\end{document}